**RedFame**

# Regional Inequality Simulations Based on Asset Exchange Models with Exchange Range and Local Support Bias

Takeshi Kato[1], Yasuyuki Kudo[1], Hiroyuki Mizuno[2] & Yoshinori Hiroi[3]

[1] Hitachi Kyoto University Laboratory, Open Innovation Institute, Kyoto University, Kyoto, Japan

[2] Center for Exploratory Research, Research & Development Group, Hitachi, Ltd., Tokyo, Japan

[3] Kokoro Research Center, Kyoto University, Kyoto, Japan

Correspondence: Takeshi Kato, Hitachi Kyoto University Laboratory, Open Innovation Institute, Kyoto University, Kyoto 606-8501, Japan.

## Abstract

To gain insights into the problem of regional inequality, we proposed new regional asset exchange models based on existing kinetic income-exchange models in economic physics. We did this by setting the spatial exchange range and adding bias to asset fraction probability in equivalent exchanges. Simulations of asset distribution and Gini coefficients showed that suppressing regional inequality requires, firstly an increase in the intra-regional economic circulation rate, and secondly the narrowing down of the exchange range (inter-regional economic zone). However, avoiding over-concentration of assets due to repeat exchanges requires adding a third measure; the local support bias (distribution norm). A comprehensive solution incorporating these three measures enabled shifting the asset distribution from over-concentration to exponential distribution and eventually approaching the normal distribution, reducing the Gini coefficient further. Going forward, we will expand these models by setting production capacity based on assets, path dependency on two-dimensional space, bias according to disparity, and verify measures to reduce regional inequality in actual communities.

## 1. Introduction

Income inequalities and disparities across populations lead to social problems in different countries around the world. Regional inequalities within a country, i.e. inequalities between cities and regions and between regional blocks, present additional major problems. Not only do regional inequalities impact on income disparities, they also lead to inequalities in the quality of life, such as access to employment, education, and healthcare services. Population concentration in cities in developing and developed countries leads to a vicious cycle of population outflow from villages into urban areas, an increase in price levels of housing and education in cities, a decrease in birth rate, a decline in the national population, a shift of employment opportunities and medical and educational services towards urban areas which further propagates the population concentration in cities with aging populations left in rural areas.

According to the latest OECD report (OECD, 2018), since 2000, regional inequality within half of the OECD countries has either leveled off or decreased, however, inequalities in the other OECD countries have increased. Income and employment opportunities are still concentrated in large cities and specific regions. Creating a sustainable and inclusive society means that inequalities due to spatial disparity ought to be addressed. The OECD's declaration against regional inequality, i.e. spatial inequality, is consistent with the UN's SDG 10 to reduce inequality within and among countries and goal 11 to make cities inclusive, safe, resilient, and sustainable (United Nations, 2015).

The Zipf's law, a well-known statistical empirical law in spatial economics for cities (Zipf, 1949), states that a city's size (i.e. its' population or economic scale) and its ordinal rank follow a certain power law. Recently, the Zipf's law has been used in the OECD to examine the relationship between a region's size (regional GDP) and the heterogeneity its' growth. This has been done by comparing the distribution of regional contribution ratio to GDP growth and the distribution based on Zipf's law (Garcilazo & Martins, 2015). Zipf's law has been used to illustrate that due to the coherence of culture within the province, the province-level size distribution of Pakistani cities at the national and provincial level follows the Zipf's law (Arshad, Hub & Ashraf, 2019). Additionally, the distribution of city size in China, Nigeria, and India are based on urbanization policies, historical backgrounds and urban development (Farrell & Nijkamp, 2019).

One research study examined the city size using the perspective of economic physics in relation to Zipf's law (Ghosh, Chatterjee, A. S. Chakrabarti, & B. K. Chakrabarti, 2014). In this study, the growth of cities was studied as a "Kolkata Paise Restaurant" problem; a model for calculating the distribution of customer numbers (city population) against the intrinsic fitness of restaurants (cities). Cities' fitness here referred to the available services in restaurants which for cities corresponded to wealth, employment, wages, transportation, housing, and other economic indicators. The distribution of fitness followed either a uniform distribution or a power-law distribution. Since the probability of a population staying in the same city is based on the population's satisfaction with its' services stay, unsatisfied populations randomly opt for other cities. The model robustly captured the differences in distribution of fitness and suitably explained the city size distribution, however, the authors did not delve into reasons why differences in fitness distribution, i.e. regional economic disparities, occur.

The Pareto Principle, a well-known statistical empirical law for income and wealth distribution in mathematical economics (Pareto, 1896), is a power law expressing the uneven distribution of wealth. It states that 80% of the total wealth is owned by the upper 20% of the wealthy class. Typical studies in economic physics derive the distribution of wealth from the Pareto Principle and gamma distribution. This is done using kinetic exchange models based on the interaction of two agents, as simulated by the analogy of kinetic energy exchange during two-particle ideal gas collisions (Chatterjee & Chakrabarti, 2007; Chatterjee, Chakrabarti, & Manna, 2004). In these models the total income of the two agents (excluding their savings) is randomly exchanged; these microscopic processes are repeated over and over to derive the wealth distribution as a macroscopic equilibrium. Since the model simulates particle collision, it does not consider the concepts of spatial distance and equivalent exchange in economic transactions.

Using a constructive approach in economic physics, we investigated the reasons behind the occurrence of regional inequalities by incorporating the spatial and the economic transaction perspectives which are basic to the study of regional economics. Since regional economic disparities are related to disparities in the quality of life, we assumed that Zipf's law for city size would hold true if the population distribution corresponds to the distribution of fitness as shown in the Kolkata Paise Restaurant problem. In Sections 1 and 2, this paper that focuses on regional economic disparity, presents background information on the topic. In Section 3, we propose asset exchange models for regional economic activities in reference to the above kinetic exchange models and incorporate the concepts of intra-regional economic circulation and inter-regional exchange range into the models. Moreover, we also incorporate the concept of local support from cities after considering equal exchange of assets. In Section 4, we show results of simulations of regional asset distribution based on these models using intra-regional economic circulation rate, inter-regional exchange distance, and local support bias as parameters. We also show results of the Gini coefficient, a typical measure of disparity, to elucidate regional inequalities. In Section 5, we compare these results with the Pareto Principle and the gamma distribution and propose guidelines to reduce regional inequality. Finally, in Section 6, we present our conclusions and future prospects in solving the social problem of regional inequality.

## 2. Literature Review

Pareto's work explained how wealth and income are distributed; he showed that wealth distribution follows a power law (Pareto, 1896). Champernowne later explained Pareto's law and proposed a model which showed that income distribution evolves over time using a stochastic process (Champernowne, 1953). According to a review by Yakovenko & Rosser, numerous subsequent studies have shown that the distribution of income and wealth approximates both lognormal and gamma distributions, and that the tails of these distributions follow a power law (Yakovenko & Rosser, 2009). Following this, various studies have been conducted to explain these distributions using a stochastic process and a multi-agent asset exchange model.

The surplus theory of social stratification (proposed by a sociologist Angle) is known as a representative theory of the distribution of individual wealth (Angle, 1986). This surplus theory illustrates the process of inequality caused by stochastic processes that generate gamma-like distributions to explain the tendency of wealth to flow into already wealthy people and the ability to maintain greater wealth generation. Specifically, all economic agents save a certain percentage of their wealth, and two arbitrarily selected economic agents randomly decide the amount to exchange from their surplus of wealth that excludes their savings, and randomly decide which of them will receive the exchange amount.

Hayes and Chakraborti explained how the rich become richer and the poor become poorer from the viewpoint of economic physics by regarding the economic system as a multi-body physical system (Hayes, 2002; Chakraborti, 2002). Just as the exchange of momentum due to random collisions between gas molecules causes macroscopic properties such as temperature and pressure, in this model, the random exchange of wealth between individuals in an economic system causes macroscopic phenomenon such as the distribution of wealth. Specifically, two arbitrarily selected economic agents randomly decide the exchange amount within a range that does not exceed the size of the wealth of the poor, and also randomly decide which one of them will receive the exchange amount. The difference from the Angle model is that the exchange amount is determined according to the amount the poor have, and if the exchange is repeated for a long period of time, almost all the wealth will be concentrated in one economic agent.

Several models that develop the Angle, Hayes, and Chakraborti models which change the method of determining the exchange amount have been proposed. In one model, an exponential distribution is obtained by randomly dividing the exchange amount as opposed to one of the two economic agents receiving the exchange amount (A. S. Chakrabarti & B. K. Chakrabarti, 2010). In another model, a gamma distribution is obtained by adding a specific rate of savings to all economic agents (Chatterjee & Chakrabarti, 2007); this has been was studied in detail through fitting of simulation data and proven to follow a gamma distribution (Patriarca, Chakraborti, & Kaski, 2004). Furthermore, a model in which the savings rate of economic agents follows a uniform distribution yields a power-law distribution (Chatterjee, Chakrabarti, & Manna, 2004).

Two newer typical models introduced the idea of taxation and insurance to control wealth inequality; the tax model and the insurance model. In the tax model, a fixed tax rate is levied on the wealth of two economic agents, and the total amount of tax is evenly distributed to all economic agents after randomly dividing the exchange amount (Guala, 2009). In this model, as the fixed tax rate rises, it shifts from an exponential distribution to a gamma distribution and subsequently returns to an exponential function. In the insurance model, economic agents are insured to avoid risks; exchanges are only made after the winner agrees to transfer a portion of the excess of wealth to the loser (A. S. Chakrabarti & B. K. Chakrabarti, 2009). In this model, as the transfer ratio increases, the exponential distribution shifts to the Δ distribution via the gamma distribution.

So far, we have reviewed a portion of the most representative literature in the field. In this paper we, focus on regional inequality, use wealth exchange to represent asset exchange in the regional economy, and use wealth savings to represent regional economic circulation. In attempt to portray a real-life situation as a method of exchanging assets, the amount of assets exchanged between rich and poor regions will be based on the amount that poor regions can contribute. As shown by the Hayes and Chakraborti model, this situation creates a large regional inequality. We will therefore introduce concepts such as a regional exchange range and a local support bias based on the regional economy in place of the tax exchange and the insurance.

## 3. Economic Exchange Models

### 3.1 Kinetic Exchange Models

A few kinetic exchange models in economic physics have been proposed to explain income and wealth distribution (A. S. Chakrabarti & B. K. Chakrabarti, 2010; Takita, 2012). These models are based on the exchange of kinetic energy (or momentum) that occurs during the collision of ideal gas particles. Although there are variations in these models, the typical models are the CC1 Model (A. S. Chakrabarti & B. K. Chakrabarti, 2010), the HC Model (Hayes, 2002; Chakraborti, 2002), the Angle Model (Angle, 1986), the CC2 Model (Chatterjee & Chakrabarti, 2007), and the CCM Model (Chatterjee, Chakrabarti, & Manna, 2004). The model names used here are for convenience. Table 1 below describes both the existing models and our models.

Table 1. Economic exchange models

| Model | CC1 | HC | Angle | CC2 | CCM | Our model | |
| --- | --- | --- | --- | --- | --- | --- | --- |
| | | | | | | Ranged model | Biased model |
| Saving propensity | — | — | Constants common to all agents | Constants common to all agents | Factors uniformly distributed | Constants common to all agents | Constants common to all agents |
| Exchange amount | Total amount of both agents | Twice the amount of poorer agent | Total amount of both agents excluding savings | Total amount of both agents excluding savings | Total amount of both agents excluding savings | Twice the amount of poorer agent excluding saving | Twice the amount of poorer agent excluding saving |
| Exchange distance | — | — | — | — | — | Attenuation of the amount for distance | — |
| Exchange fraction | Random fraction between both | One agent randomly chosen | One agent randomly chosen | Random fraction between both | Random fraction between both | Random fraction between both | Random fraction between both |
| Poor support bias | — | — | — | — | — | — | Fraction with bias for poorer agent |
| Steady state distribution | Exponential distribution | One agent gets all | Gamma distribution | Gamma distribution | Power-law distribution | (Described later) | (Described later) |
| References | Chakrabarti & Chakrabarti, 2010 | Hayes, 2002; Chakraborti, 2002 | Angle, 1986 | Chatterjee & Chakrabarti, 2007 | Chatterjee, Chakrabarti & Manna, 2003 | (This paper) | (This paper) |

In the CC1 Model, first, two agents, $i, j (= 1, 2, \cdots, N)$, are chosen randomly from an $N$ number of agents, and the total income of the two agents is randomly divided. The income for agent $i$ for time $t$ is expressed as $m_i(t)$, and the income for agent $j$ for time $t$ is expressed as $m_j(t)$. When the two agents $i, j$ exchange the total amount of income at a random fraction probability $\varepsilon$, their incomes at time $t + 1$, namely, $m_i(t + 1)$, $m_j(t + 1)$, are expressed by equations (1). Here, the probability $\varepsilon$, is a uniform random number defined by the range $0 \leq \varepsilon \leq 1$. Setting the initial value of incomes of the $N$ agents equally, and repeating the exchange process of the CC1 Model, leads to convergence of the income frequency distribution into a steady state, and as a result, an exponential distribution is attained.

CC1 model:
$$m_i(t + 1) = \varepsilon \cdot \left( m_i(t) + m_j(t) \right)$$
$$m_j(t + 1) = (1 - \varepsilon) \cdot \left( m_i(t) + m_j(t) \right)$$

$$(1)$$

In the HC Model, the amount of exchange $\text{Min}\left( m_i(t), m_j(t) \right)$ is decided based on the income of the poorer of the two agents $i, j$ randomly chosen from $N$ number of agents. The amount of exchange provided by both agents is allotted to one of the agents based on a random probability $\varepsilon'$. The incomes after the exchange $m_i(t + 1)$, $m_j(t + 1)$ are expressed by equations (2). Here, the probability $\varepsilon'$, is a random integer that is either $0$ or $1$. Despite having equal initial values for income, repeating the exchange process of the HC Model leads to drastic results. All the incomes concentrate to one agent, and the other agent loses all his income.

HC model:
$$m_i(t + 1) = m_i(t) - \text{Min}\left( m_i(t), m_j(t) \right) + 2 \cdot \varepsilon' \cdot \text{Min}\left( m_i(t), m_j(t) \right)$$
$$m_j(t + 1) = m_j(t) - \text{Min}\left( m_i(t), m_j(t) \right) + 2 \cdot (1 - \varepsilon') \cdot \text{Min}\left( m_i(t), m_j(t) \right)$$

$$(2)$$

The Angle Model and the CC2 Model incorporate savings propensity. The two agents $i, j$ save part of their incomes at time $t$ at a savings rate of $\lambda$. The savings rate $\lambda$ is a common fixed number for the $N$ number of agents. The Angle Model and the CC2 Model have different exchange methods. In the Angle Model, the remaining income after deducting the savings of both agents $(1 - \lambda) \cdot \left( m_i(t) + m_j(t) \right)$, is allotted to one of the agents at a random probability $\varepsilon'$ as with the HC Model. In the CC2 Model, the remaining income after deducting the savings of both agents is exchanged at a random fraction probability $\varepsilon$ as with the CC1 Model. The income after exchange in the CC2 Model $m_i(t + 1)$, $m_j(t + 1)$ is expressed by equations (3). Substituting the fraction probability $\varepsilon$ in equations (3) with a random integer $\varepsilon'$ derives the equations for expressing the Angle Model. Repeating the exchange process of the Angle Model and CC2 Model enables attaining a gamma distribution with a smaller income disparity compared to that of the CC1 Model and HC Model. Increasing the savings rate $\lambda$ further enables suppressing income disparity to a higher degree. Due to the difference in probability $\varepsilon'$ and probability $\varepsilon$, the CC2 Model suppresses income disparity better than the Angle Model.

CC2 model:
$$m_i(t + 1) = \lambda \cdot m_i(t) + \varepsilon \cdot (1 - \lambda) \cdot \left( m_i(t) + m_j(t) \right)$$
$$m_j(t + 1) = \lambda \cdot m_j(t) + (1 - \varepsilon) \cdot (1 - \lambda) \cdot \left( m_i(t) + m_j(t) \right)$$

$$(3)$$

The CCM Model has a different savings method than that of the CC2 Model. It deals with cases wherein the savings rate differs for each of the $N$ number of agents. Substituting the savings rate $\lambda$ in equations (3) with savings rates $\lambda_i, \lambda_j$ for agents $i, j$ enables deriving the equations for expressing the CCM Model. The savings rates $\lambda_i, \lambda_j$ have a uniform distribution defined by the range $0 \leq \lambda_i, \lambda_j \leq 1$, which remain the same between exchange processes. Repeating the exchange process of the CCM Model enables attaining a power-law distribution that follows the Pareto Principle. Due to the difference between the common savings rate $\lambda$ and the variable savings rates $\lambda_i, \lambda_j$, the power-law distribution of the CCM Model leads to a larger income disparity than the gamma distribution of the CC2 Model. In other words, with the same savings rate, there is better suppression of income disparity.

### 3.2 Regional Asset Exchange Models

We revised the existing kinetic models based on regional economic activities to examine regional inequality. Income exchange in the existing models represents the exchange of assets in the regional economy. Savings from income represents the channelling of assets back into the region, i.e. an intra-regional economic circulation. In the CC2 Model, since only the income left over after savings are deducted is exchanged between two agents $i, j$, the richer agent contributes a larger amount than the poorer agent. However, in view of the actual situation of asset exchange it is more appropriate to allocate an equal amount of exchange to the two agents $i, j$, based on the remaining income that the poorer agent is able to contribute, as with the HC Model.

Firstly, we came up with a compromise between the HC Model and the CC2 Model on the basis of the regional asset exchange models. Assets for region $i$ at time $t$ is represented as $m_i(t)$, and assets for region $j$ at time $t$ is represented as $m_j(t)$. For the two regions $i, j$, part of the assets is channelled back into the region at an intra-regional economic circulation rate $\lambda$ at time $t$. Of the remaining assets left over after deducting intra-regional circulation for the two regions $i, j$, the remaining assets of the poorer region $(1 - \lambda) \cdot \text{Min} \left( m_i(t), m_j(t) \right)$ are set as the exchange amount for each of the two regions $i, j$. When assets are exchanged at a random fraction probability $\varepsilon$ between the two regions $i, j$, each of the assets at time $t + 1$, i.e. $m_i(t + 1)$, $m_j(t + 1)$, can be expressed by equations (4) (hereinafter referred to as the "Base Model"). The fraction probability $\varepsilon$ is a uniform random number defined by the range $0 \leq \varepsilon \leq 1$. To understand the trend for all the regions, we set the intra-regional economic circulation rate $\lambda$ as a fixed number for all regions; although it is possible to have different intra-regional economic circulation rates for the regions of $N$ as with the CCM Model.

Base Model:

$$m_i(t + 1) = m_i(t) - (1 - \lambda) \cdot \text{Min} \left( m_i(t), m_j(t) \right) + 2 \cdot \varepsilon \cdot (1 - \lambda) \cdot \text{Min} \left( m_i(t), m_j(t) \right)$$
$$m_j(t + 1) = m_j(t) - (1 - \lambda) \cdot \text{Min} \left( m_i(t), m_j(t) \right) + 2 \cdot (1 - \varepsilon) \cdot (1 - \lambda) \cdot \text{Min} \left( m_i(t), m_j(t) \right)$$

$$(4)$$

Secondly, we incorporated the spatial perspective into the Base Model shown equations (4). The amount of exchange varied based on the distance between two regions; it was high between neighboring regions and low between distant regions. We also took into consideration the economic zone between neighboring regions, i.e. the inter-regional zone. The regions of N were arranged in numerical order at a distance interval of $1$ along a one-dimensional axis; the distance range $\omega$ for possible asset exchange between the two regions $i, j$ was defined, as shown in Figure 1. Initially, the first region $i$ was randomly selected among the regions within the range from $1 + \omega$ to $N - \omega$ along the one-dimensional axis, and the second region $j$ was selected within the $i \pm \omega$ range. The distance between the two regions $i, j$ was expressed as $x_{ij}$. The amount of exchange in each region $i, j$ was the value obtained by multiplying the logistic distribution function $f(x_{ij}; 0, \sigma)$ with the amount of exchange expressed in equations (4) $(1 - \lambda) \cdot \text{Min} \left( m_i(t), m_j(t) \right)$. The logistic distribution is widely seen in populations of living organisms and in the dispersion of energy resources. Herein, as $f(x_{ij}; 0, \sigma)$, we used the average $0$ and the normalized probability density function of dispersion $\pi^2 \sigma^2 / 3$, and set $\sigma = \omega / 10$. As such, the regional asset exchange model that incorporated the spatial concept was expressed by equations (5) (hereinafter referred to as the "Ranged Model").

Ranged Model:

$$m_i(t + 1) = m_i(t) + (2 \cdot \varepsilon - 1) \cdot (1 - \lambda) \cdot \text{Min} \left( m_i(t), m_j(t) \right) \cdot f(x_{ij}; 0, \sigma)$$
$$m_j(t + 1) = m_j(t) + (2 \cdot (1 - \varepsilon) - 1) \cdot (1 - \lambda) \cdot \text{Min} \left( m_i(t), m_j(t) \right) \cdot f(x_{ij}; 0, \sigma)$$

$$f(x_{ij}; 0, \sigma) = \frac{4 \cdot \exp \left( -\frac{x_{ij}}{\sigma} \right)}{\left( 1 + \exp \left( -\frac{x_{ij}}{\sigma} \right) \right)^2}$$

$$(5)$$

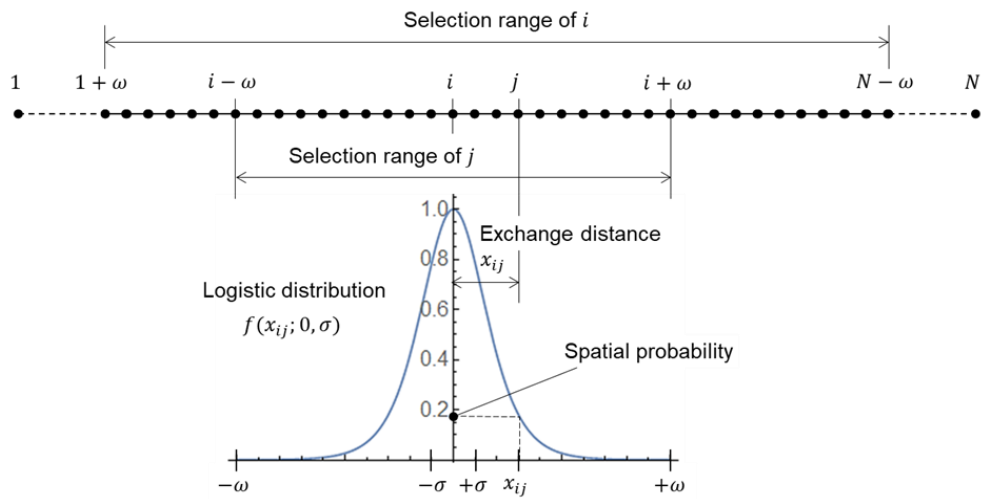

Figure 1. Incorporating spatial concepts in regional asset exchange model

In the Ranged Model, the asset exchange amount $(1-\lambda) \cdot \text{Min}\left(m_i(t), m_j(t)\right) \cdot f\left(x_{ij}; 0, \sigma\right)$ decreased with increasing the distance $x_{ij}$ between regions $i, j$. In the same way that income disparity could be suppressed better in the CC2 Model by incorporating the savings rate $\lambda$ and decreasing the amount of exchange when compared to the CC1 Model, it would also be possible to reduce regional inequality in the Ranged Model as shown in equations (5) and the Base Model shown in equations (4). Another possible approach to reduce regional inequality would be to provide support to the poorer regions in the exchange of assets. In the context of a region's economic activities, this would amount to richer cities providing support to the regions. The fraction probability $\varepsilon$ shown in equations (4) was substituted with the fraction probability $\varepsilon''$, which included local support bias. This kind of bias could be considered as an on-going distribution norm that was embedded in the exchange process, rather than an ex-post asset redistribution. In particular, a probability bias $b$ $(0 \le b \le 1)$ was assigned beforehand to the poorer region, and the amount of exchange $(1-\lambda) \cdot \text{Min}\left(m_i(t), m_j(t)\right)$ was randomly divided at a probability of $(1-b) \cdot \varepsilon$, where $\varepsilon$ was a uniform random number defined by the range $0 \le \varepsilon \le 1$. As such, the regional asset exchange model that incorporated the concept of support for the poorer region was described using equations (6) (hereinafter referred to as the "Biased Model").

Biased Model:

$$m_i(t+1) = m_i(t) + \left(2 \cdot \varepsilon'' - 1\right) \cdot (1-\lambda) \cdot \text{Min}\left(m_i(t), m_j(t)\right)$$
$$m_j(t+1) = m_j(t) + \left(2 \cdot \left(1 - \varepsilon''\right) - 1\right) \cdot (1-\lambda) \cdot \text{Min}\left(m_i(t), m_j(t)\right)$$
$$\quad if \ m_i(t) < m_j(t) \quad \varepsilon'' = b + (1-b) \cdot \varepsilon$$
$$\quad if \ m_i(t) = m_j(t) \quad \varepsilon'' = \varepsilon$$
$$\quad if \ m_i(t) > m_j(t) \quad \varepsilon'' = 1 - b - (1-b) \cdot \varepsilon$$

$$(6)$$

The 'taxation and redistribution' model (Guala, 2009) and the 'risk aversion and insurance' model (A. S. Chakrabarti & B. K. Chakrabarti, 2009) are known models that are similar to the Biased Model. In the former model (i.e. the taxation model), tax from two agents, is collected at the same rate and evenly distributed between the two agents. In the latter model (i.e. the insurance model), the excess of wealth between the winner and the loser in the exchange is multiplied by a pre-agreed transfer ratio to, and the insurance amount is collected from the winner and transferred to the loser. Since both models are based on the same idea as the CC2 Model, the rich agent contributes a larger exchange amount than the poor agent. In order to examine regional inequality, we believed that the bias effect would be better clarified if support for the poor agent (as with the Biased Model) was added to the fraction probability explicitly after setting an equivalent exchange condition, as was done with our Base Model.

The major features that distinguish our models from the existing models was the inclusion of the inter-regional exchange range from the Ranged Model and the local support bias from the Biased Model, as shown in Table 1. Moreover, by comparing equations (5) and (6), the difference between the Ranged Model and the Biased Model is the presence/absence of the logistic distribution function $f\left(x_{ij}; 0, \sigma\right)$ and the difference between the fraction probabilities $\varepsilon$ and $\varepsilon''$. A model combining the $f\left(x_{ij}; 0, \sigma\right)$ and $\varepsilon''$ (hereinafter referred to as the "Ranged-Biased Model") can easily be conceived, as shown in equations (7). The definitions of the spatial probability $f\left(x_{ij}; 0, \sigma\right)$ in the exchange range $\omega$ and the fraction probability $\varepsilon''$ with local support bias $b$ in equations (7) are the same as with equations (5) and (6), respectively.

Ranged-Biased Model

$$m_i(t+1) = m_i(t) + \left(2 \cdot \varepsilon''' - 1\right) \cdot (1-\lambda) \cdot \text{Min}\left(m_i(t), m_j(t)\right) \cdot f(x_{ij}; 0, \sigma)$$

$$m_j(t+1) = m_j(t) + \left(2 \cdot \left(1 - \varepsilon'''\right) - 1\right) \cdot (1-\lambda) \cdot \text{Min}\left(m_i(t), m_j(t)\right) \cdot f(x_{ij}; 0, \sigma)$$

$$(7)$$

## 4. Asset Distribution and Gini Coefficient Simulations

Herein, we describe the numerical simulations pertaining to the asset distribution for the Ranged Model, the Biased Model, and the Ranged-Biased Model. The majority of regions have a hierarchical structure; the number of countries in the world, provinces in a country, cities in a province, and towns in a city range from 10 to 100. Since exchange between regions may take place across different hierarchies, in our simulations, we set the number of regions to $N = 1000$. The initial value of assets was equally set for all regions to 1 ($m_i(0) = 1, i = 1, 2, \cdots, N$). For the intra-regional economic circulation rate, three cases of $\lambda = 0, 0.4, 0.8$ were set as parameters. For the inter-regional zone exchange range, a sequence of regions $1 \sim N$ along a one-dimensional axis was assumed; three cases of $\omega = 3, 10, 30$ were set as parameters. For the local support bias expressing the distribution norm, three case of $b = 0, 0.2, 0.4$ were set as parameters, including a case where there was no bias. The exchange repeat count was set to a sufficiently large frequency of $t_{max} = 10^5$ to achieve an asset distribution that approached the steady state.

The results of simulation of regional asset distribution for the existing CC2 Model and our Base Model are shown in Figure 2. In the CC2 Model, although the frequency distribution of asset $m$ became an exponentially-shaped distribution at an intra-regional economic circulation rate of $\lambda = 0$, it became a normally-shaped distribution with a mean value of 1 as the initial asset value, as the rate increased from $\lambda = 0.4 \rightarrow 0.8$. Conversely, in the Base Model, as with the HC Model, assets concentrated within a few regions at $\lambda = 0$, with other regions losing almost all their assets. Although this tendency eased as $\lambda = 0.4 \rightarrow 0.8$, a further increase in the exchange repeat count $t_{max}$ eventually led to all assets concentrating in one region (explained later in Figure 6).

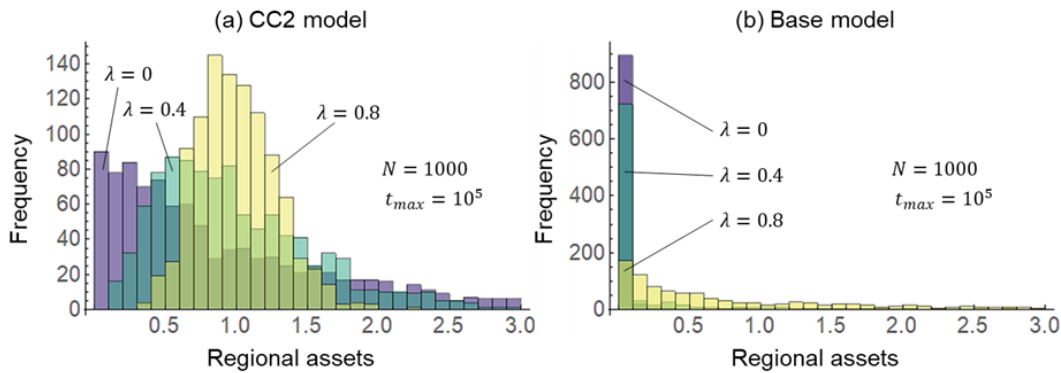

Figure 2. Regional asset distributions of (a) CC2 model and (b) our Base Model

Figure 3 shows the results of simulation of asset distribution for the Ranged Model and the Biased Model in the case of an intra-regional economic circulation rate of $\lambda = 0.4$. In the Ranged Model, as the exchange range became narrower from $\omega = 30 \rightarrow 3$, the asset distribution moved from an exponentially-shaped distribution and approached a normally-shaped distribution. In the Biased Model where $b = 0$, the case was the similar to that of the Base Model; assets tended to concentrate within a few regions. As the bias increased from $b = 0.2 \rightarrow 0.4$, the asset distribution from an exponentially-shaped distribution approached a normally-shaped distribution similar to that of the Ranged Model. However, this may be a little hard to visualize as the frequency range was adjusted to that when the case $b = 0$.

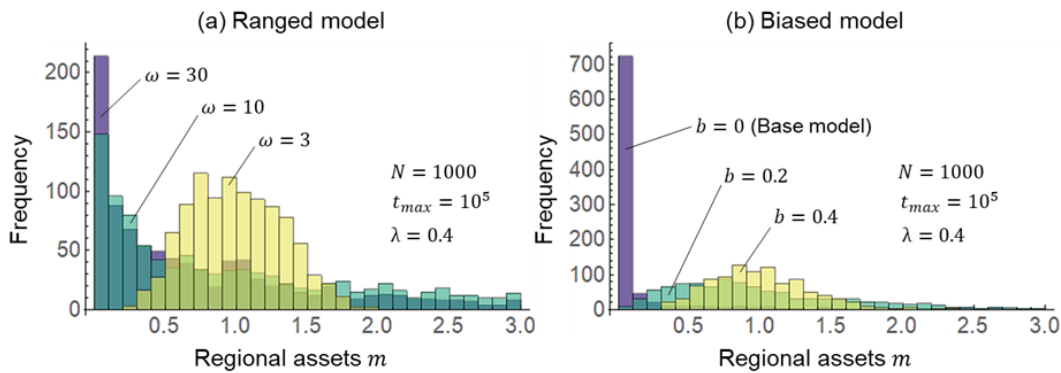

Figure 3. Regional asset distributions of (a) Ranged Model and (b) Biased Model

We then computed the Gini coefficient to elucidate regional inequalities. Gini coefficient, a typical measure of disparity of incomes, can also be applied to regional inequality. The Gini coefficient, obtained from the frequency distribution, is computed by arranging the regions starting with the region with the smallest to the largest assets, drawing the Lorenz curve which plots the regions on the horizontal axis and the total value of assets on the vertical axis, drawing the equal distribution line that represents the case when there is absolutely no disparity in assets, and doubling the area covered by both the curve and the line. For example, the Gini coefficient for initial distribution of assets is $0$, while that for converged distributions of the HC Model or the Base Model is $1$.

Figure 4 shows a 3D graph of the Gini coefficients computed for the Ranged Model and the Biased Model. In the Ranged Model, the Gini coefficient eventually approached $0$ as the economic circulation rate $\lambda$ increased and as the exchange range $\omega$ decreased where regional inequality was suppressed. Since inter-regional exchange was negligible at $\omega = 1$, the Gini coefficient was almost $0$. Even in the Biased Model, despite having a 3D surface topology that was somewhat different from that in the Ranged Model, the Gini coefficient eventually approached $0$ as the circulation rate $\lambda$ and bias $b$ became larger.

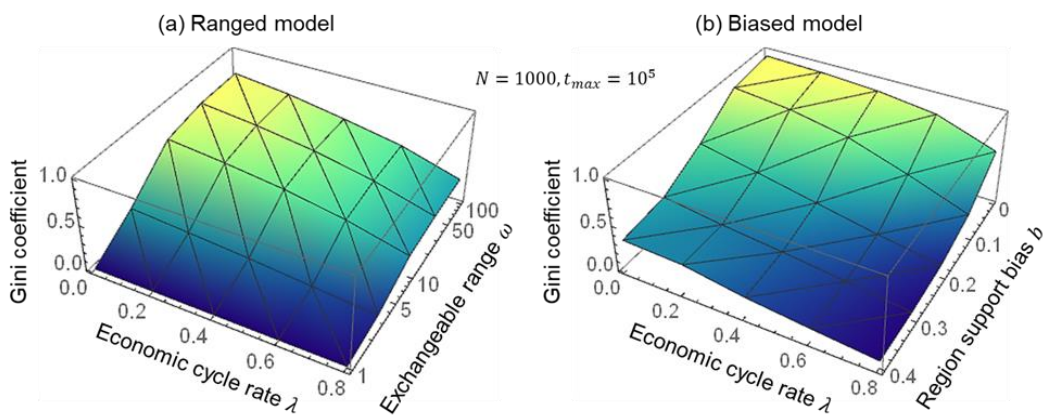

Figure 4. Gini coefficients of (a) Ranged Model and (b) Biased Model

Figure 5 shows the results of simulation of asset distribution and Gini coefficient based on the Ranged-Biased Model, a combination of the Ranged Model and the Biased Model. An intra-regional economic circulation rate of $\lambda = 0.4$ was set as common parameter. Although the asset distribution became an exponentially-shaped distribution at $\omega = 30, b = 0$, it was a normally-shaped distribution with a mean value of 1 as the initial asset value at $\omega = 10, b = 0.2$, and a delta-shaped distribution where the assets concentrate in the vicinity of $1$ at $\omega = 3, b = 0.4$. The Gini coefficient, approximately $0.6$ at an exchange range of $\omega = 100$ and bias of $b = 0$, became smaller as the exchange range became narrower and bias became larger, eventually approximating $0$ at the vicinity of $\omega = 1$ and $b = 0.4$.

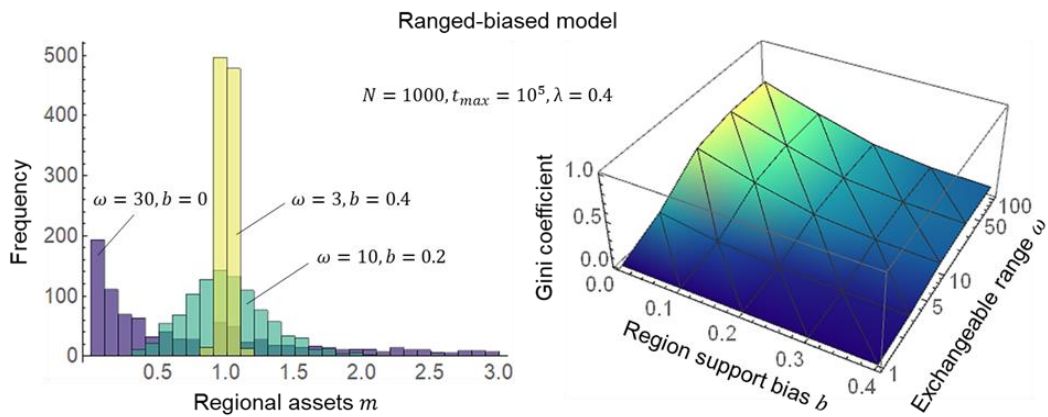

Figure 5. Regional asset distribution and Gini coefficients of Ranged-Biased Model

Figure 6 shows results of calculation of Gini coefficient using the exchange repeat count $t$ as a variable of repeat counts. In the CC2 Model, the Gini coefficient converged to the vicinity of $0.3$ from a repetition frequency of $t = 10^4$. In the Base Model, as with the HC Model, the Gini coefficient eventually approached $1$, and all the assets concentrated in one region. In the Ranged Model, the Gini coefficient was lesser than that of the Biased Model when the value of $t$ was small, but it eventually approached $1$ as $t$ increased. In the Biased Model, the Gini coefficient converged to between $0.3 \sim 0.4$ after $t = 10^4$. The Ranged-Biased Model exhibited a Gini coefficient similar to the Ranged Model when the value of $t$ was small but converged to a value lower than $0.2$ after $t = 10^4$. In summary, the Gini coefficient converged to $1$ in the Base Model and the Ranged Model, while it converged to a relatively smaller value in the CC2 Model, the Biased Model, and the Ranged-Biased Model. Among the later three, the Ranged-Biased Model had the lowest Gini coefficient; therefore, regional inequality was suppressed.

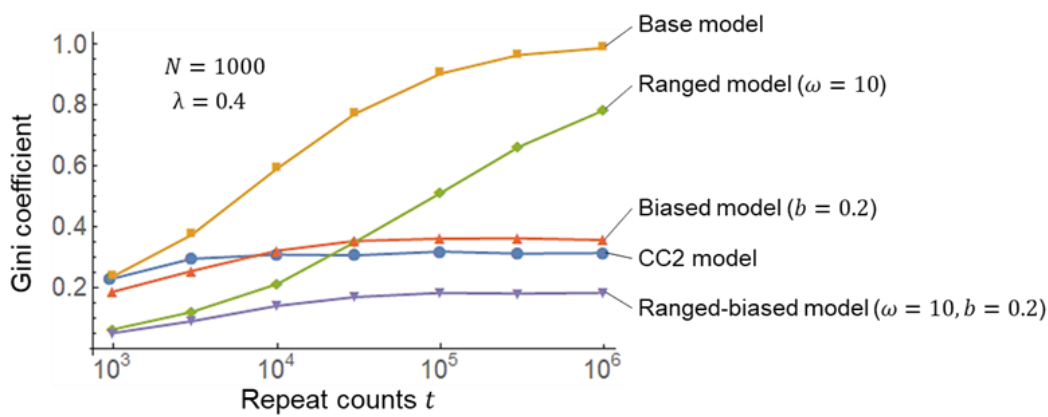

Figure 6. Gini coefficients of regional asset exchange models for exchange repeat counts

To vividly illustrate regional inequality, we created a network graphs to depict asset distribution in the Base Model, the Ranged Model, the Biased Model, and the Ranged-Biased Model. The size of the network vertices represents the size of the regional assets and the edges represent the history of asset exchange. Here we set the number of regions to $N = 100$ and the repeat count to $t_{max} = 2000$, with an intra-regional economic circulation rate of $\lambda = 0.4$ as a common parameter. The Base Model and the Biased model had no spatial dimension, and the Ranged Model and the Ranged-Biased Model had a one-dimensional space. In order to compare the Ranged Model and the Ranged-Biased Models, the network graph was drawn on a two-dimensional plane using a spring model. In the Base Model and the Biased Model, the inter-regional networks were spread on a two-dimensional plane, however, in the Ranged Model and the Ranged-Biased Model, the networks were arranged in what approximated a one-dimensional axis. There was a concentration of assets in certain regions in the Base Model, which was suppressed by applying local support bias $b = 0.2$ in the Biased Model. In the Ranged Model, since the exchange range was restricted to $\omega = 20$, asset concentration did not readily take place. In the Ranged-Biased Model, the synergistic effect between exchange range $\omega$ and bias $b$ led to a dispersion of assets.

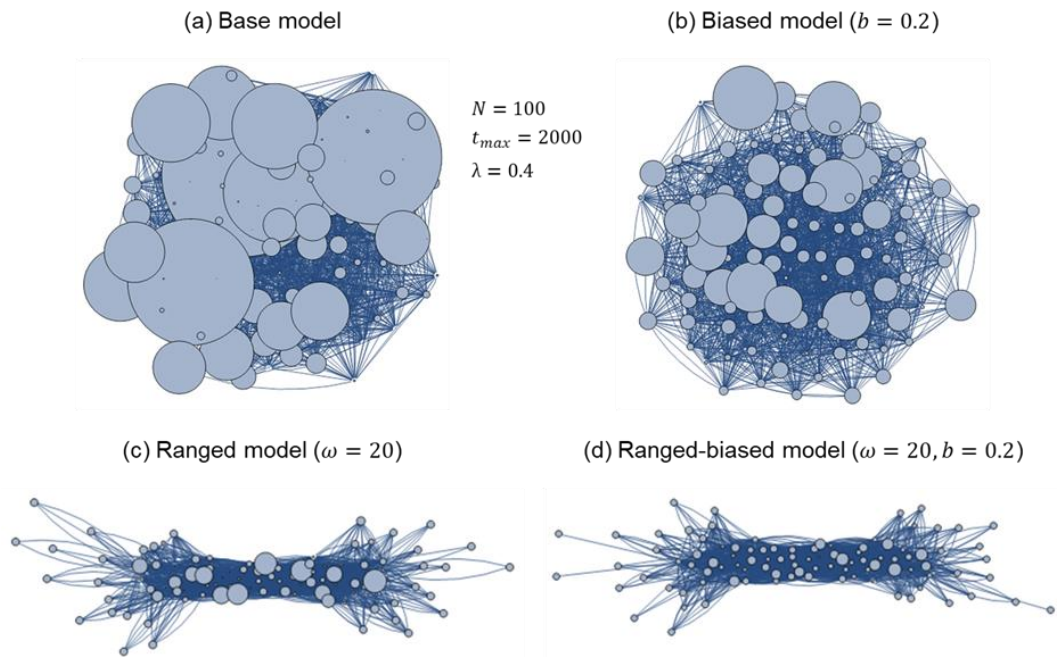

Figure 7. Network Graphs of (a) Base, (b) Biased, (c) Ranged and (d) Ranged-Biased Models

## 5. Discussions

### 5.1 Review of Results

We illustrated the results of simulations of regional asset distribution using asset exchange models. We saw how the asset distribution varied between the Base Model, the Ranged Model, the Biased Model, and the Ranged-Biased Model from a drastic distribution in which assets concentrated in one region, to an exponential distribution in which the concentration of assets was eased into a normal distribution with a mean of 1, and to a distribution that remained within the vicinity of asset initial value of 1, based on changes in parameters of intra-regional economic circulation rate $\lambda$, exchange range $\omega$, and local support bias $b$. The existing CC2 Model demonstrated that income distribution followed a gamma distribution that was based on a savings rate $\lambda$ parameter; the gamma-shaped distribution shifted from an exponential to a normal distribution as $\lambda$ increased, based on an analogy of the exchange of kinetic energy (or momentum) during the collision of gaseous particles (Patriarca, Chakraborti, & Kaski, 2004). Since our asset exchange models adjusted the amount of exchange based on the poorer region and also incorporated parameters such as exchange range $\omega$ and bias $b$, it was difficult to find mathematical analytic solutions analogous to those provided by particle collision. Nevertheless, it seemed appropriate to assume that the distributions approximated the gamma distribution.

As regards the Gini coefficient, which expresses regional inequality, assets concentrated in one region and the Gini coefficient approached 1 as the exchange repeat count $t$ increased in the Base Model and the Ranged Model. This was because our models assumed a realistic situation, i.e. an equivalent exchange of assets among regions. In the existing CC2 Model, the richer region contributed a larger amount of exchange than the poorer region, and the exchanged amount was randomly divided between the regions, the poorer region had the opportunity to recover its assets. However, in the Base Model and the Ranged Model, which were patterned after the HC Model, the amount of exchange with the richer cities was decided equally based on the assets of the poorer regions, therefore, the poorer regions lost an opportunity to recover their assets and eventually lost them all. In contrast to the Base Model and the Ranged Model, the Gini coefficients converged to a relatively smaller value without approaching 1, through the addition of local support bias $b$ in the Biased Model and the Ranged-Biased Model. This implies is that exchanges that enable even a small advantage for the poorer region suppress an increase in the Gini coefficient and prevent over-concentration.

As regards the inter-regional networks, in the Base Model, there was a mixture of vertices that were extremely large and vertices that were almost negligible in size due to the large regional inequality. The difference in sizes of vertices became smaller in the Biased Model and the Ranged Model, and all the vertices were similar size in the Ranged-Biased Model. Although the edges were drawn with the same thickness for all exchange histories in Figure 7, changing the edge thickness based on the amount of exchange enabled us to classify the network structure into three main types namely; (A) centralized, (B) Decentralized, and (C) Distributed (Figure 8) (Baran, 1964). From the results of simulation of asset distribution using our models, the concentration of assets in one region fell under the Centralized network (A), the exponential and normal distributions fell under the Decentralized network (B), and the distribution around the vicinity of asset initial value of 1 fell under the Distributed network (C). Considering that the organic organizations of living organisms usually exhibit the Decentralized network (B), inter-regional networks are also expected to exhibit a decentralized network shape.

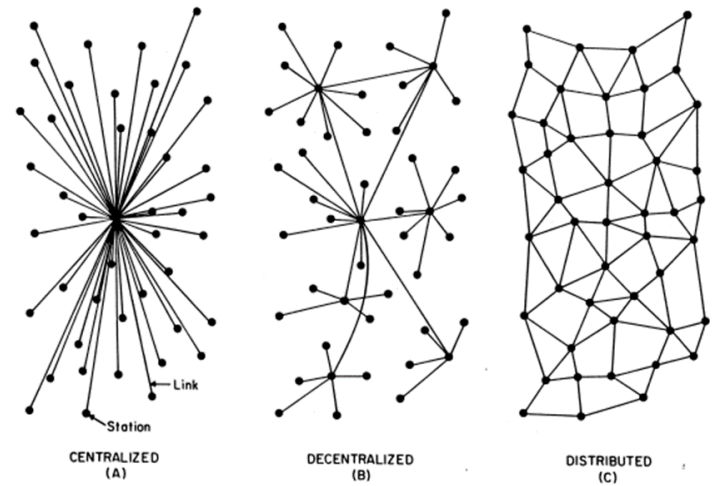

Figure 8. Network structure (Baran, 1964)

Our models only took into consideration asset exchange with reference to existing kinetic models. Although we incorporated the spatial and economic transaction perspectives, which were basic to understanding regional inequality, there is a need to model resource utilization capacity, production capacity, and asset management capacity based on regional assets, as well as the profits they generate, in order to further enhance these models. Where capacities and profits exceed the proportion of assets, it is possible to generate a more uneven distribution of assets than shown by the results of the simulations we conducted which would further increasing regional inequality. In particular, the "rate of return on capital > growth rate of income per capita" relationship demonstrated by Piketty would work towards further increasing inequality (Piketty, 2014). Furthermore, assuming a certain distribution rather than setting a fixed intra-regional economic circulation rate $\lambda$ for all the regions would most likely lead to an asset distribution with an even larger regional inequality, as manifested by the power-law distribution (Pareto Principle) in the CCM Model (when compared to the exponential distribution in the CC2 Model).

To incorporate a spatial perspective into the modelling process, we set the exchange range $\omega$ by arranging the regions along a one-dimensional axis, and set the exchange probability based on the logistic distribution function according to the exchange distance. A more realistic model can be achieved by arranging regions on a two-dimensional mesh, or by setting the exchange range $\omega$ and arranging regions according to their actual geographical location. Additionally, instead of a logistic distribution, it is possible to consider not only distance but also path dependence, as well as assign specific weights for each path. The former corresponds to the means for transporting assets and resources, while the latter corresponds to the costs of and the time used for transportation. As regards the local support bias $b$, although a fixed bias was set for all exchanges in this study, a certain distribution for bias could be set, or the bias could be altered based on the ratio of assets within regions. It would also be necessary to determine the process of providing local support as the distribution norm, to incorporate the increase in regional inequality arising from differences in production capacity and rate of return on capital that was previously mentioned.

### 5.2 Considerations

What would be the ideal distribution of regional assets and Gini coefficients? The power laws such as the Zipf's law for city size distribution and the Pareto Principle for income distribution come into play. In social networks, the scale-free property, i.e., the power law for degree distribution, is widely observed. Moreover, while body weight and labor productivity follow a log-normal distribution similar to that of a power-law distribution, human body height and academic ability as well as process capabilities of manufacturing plants are generally known to follow a standard normal distribution. Leaving a region's population or economic scale to follow the power law however, does not solve the problems of concentration in cities and regional inequality. Achieving a standard normal distribution for the distribution of regional assets requires the implementation of certain intervention measures to the asset exchange processes that cause regional inequality.

As demonstrated by the calculation of Gini coefficients shown in Figure 4, initial measures to increase the intra-regional economic circulation rate $\lambda$ are necessary. Allowing assets to flow back into the region prevents the outflow of assets from the region. Furthermore, if it is difficult to increase the economic circulation rate $\lambda$ within a region, an effective measure would be to narrow down the asset exchange range $\omega$. Setting an economic zone between neighboring regions, i.e. an inter-regional zone of range $\omega$, increases the economic circulation within the zone and complements the economic circulation within regions. In addition, considering the energy consumption volume reported by the International Energy Agency, approximately 27% of the world's resources, 31% of that of OECD countries, and 24% of that of Japan are spent on the transport of goods (IEA, 2017). Promoting economic circulation within regions and within inter-regional zones reduces the energy used for transport and contributes to efforts that prevent global warming.

The third measure that increases the local support is the bias $b$. As discussed in Section 5.1, although regional assets are equally exchanged, it is necessary to provide a favorable bias towards poorer regions. As shown in Figure 6, adding a local support bias, in addition to setting an intra-regional economic circulation rate $\lambda$ and an inter-regional zone exchange range $\omega$, minimizes the Gini coefficient. Depending on only the local support bias $b$ could diminish economic transactions between cities and regions and lead to loss of opportunities for the regions to recover their assets. Furthermore, an over-dependence on local support would instead lead to a shrinking of the regional economy. Therefore, to prevent these adverse effects, local support must be provided after revitalizing the intra-regional economic circulation and the inter-regional zone.

Although ex-post redistribution measures, such as ODAs at the international level and local grants at the national level are often carried out, we believe it is possible to define new distribution rules (public pricing) that mandate cities to provide support funds to regions in economic transactions between cities and regions as an ex-ante or on-going measure. Although regions can impose tariff on imports from outside the region and reverse tariff on exports to outside the region as a self-protection measure, this could lead to detrimental effects from retaliatory measures imposed by other regions. For example, one idea might be to have a positive measure in which cities themselves cash back tariffs to the regions when they (cities) import from the regions.

Thinking of the entire regional network as a single living organism, we can see from Figure 7 that localized enlargement of certain parts of the body and shrinkage in other parts is inimical to the health of the entire organism and would eventually lead to its death. There is a need to aim for transformation from a global economy to a regional economy so as to enable nutrients to reach all parts of the living organism. This necessitates advocacy for local foods, local production for local consumption, renewable energy, and local cultures (e.g. Norberg-Hodge, 2016; Edahiro, 2018). It also advocates for intra-regional economic circulation with the intent for an economy based on commons, communities, and cooperatives (Hardt & Negri, 2000; Graeber, 2004; Hiroi, 2009). In addition to revitalizing the economy of local communities, this study revealed that it is important to complement economic circulation beyond the region through the inter-regional zone and introduce a distribution norm for supporting the regions.

## 6. Summary

### 6.1 Conclusions

To gain insights into how suppressing regional inequality can be realized, we proposed new asset exchange models that incorporated the concepts of spatial exchange range (economic zone between neighboring regions) and probabilistic bias (economic transaction norm) into existing models. These concepts were missing from existing kinetic income-exchange models in economic physics. Based on these models, we then conducted simulations of regional asset distributions and Gini coefficients. We found that in addition to intra-regional economic circulation, an effective approach to address regional inequality could be attained through the inter-regional economy for the inter-regional zone and through the distribution norm for local support.

1) Firstly, to model regional economic activities, we proposed a Base Model that equally exchanged assets at a random fraction probability between regions after deducting the intra-regional economic circulation component ($\lambda$). Following this, we set an exchange range $\omega$ between two regions and presented a Ranged Model wherein the amount of asset exchange was determined by a logistic distribution function of the distance between two regions. Moreover, we set a bias $b$ for the fraction probability and presented a Biased Model in which asset exchange was carried out in favor of the poorer regions. The Ranged-Biased Model was a combination of the two models.

2) On the basis of the three asset exchange models, we conducted simulations of asset distributions and Gini coefficients using the following parameters: intra-regional economic circulation rate $\lambda$, exchange range $\omega$, and local support bias $b$. In the Base Model, assets concentrated in one region and the Gini coefficient converged to 1. In the Ranged Model, as the exchange range $\omega$ became narrower, the asset distribution shifted from an exponentially-shaped distribution to a normally-shaped distribution, and the Gini coefficient became smaller. In the Biased Model, as the bias $b$ became larger, the asset distribution approached a normally-shaped distribution similar to that of the Ranged Model, and the Gini coefficient became smaller. In the Ranged-Biased Model, assets were dispersed further through the synergistic effect of exchange range $\omega$ and bias $b$ in the Ranged-Biased Model.

3) From the results of the Base Model, overconcentration in cities would continue for as long as equivalent exchange is continued between regions. This can be effectively prevented by firstly implementing a measure to increase the intra-regional economic circulation rate $\lambda$ and secondly, by implementing a measure to narrow down the exchange range $\omega$ to complement the first measure; i.e., by establishing an inter-regional economic zone. However, repeating the exchange eventually leads to overconcentration and an asymptotic expansion of regional inequality. To prevent this from happening, there is a need to implement a third measure to add a slight local support bias $b$ to the equivalent exchange of assets. In other words, a new distribution norm should be introduced between rich cities and poor regions. Rather than depending on only one measure, we believe that it is imperative to implement a comprehensive scheme incorporating these three measures.

*6.2 Future Prospects*

Although the asset exchange models presented here are still underdeveloped, they provided basic insights into how regional inequality can be suppressed. We believe that more insights can be gained by considering the perspectives of production capacity and management capacity based on assets in addition to asset exchange. This is because such factors are likely lead to an increase in regional inequality and may therefore require new measures other than those presented here. Moreover, new insights may be gained by arranging geographical regions on a two-dimensional plane, considering path dependency in addition to exchange distance, and adding weights based on inequalities to the local support bias.

Continuous equal exchange of assets between regions will propagate the increase in regional inequality. Although ex-post redistribution measures such as local grants and ODAs are helpful, in the context of international relations, free trade does not eliminate regional inequality. We believe that new measures that provide on-going support are needed. However, these measures contradict existing norms and systems, and it would take time to reach a consensus. Until then, the only thing we could probably do, is to attempt to reduce the speed of the expansion of regional inequality by establishing an inter-regional zone ecosystem that stimulates as well complements the local economic circulation within the region. By conducting mathematical investigations such as this study, and by socially implementing the results of such studies in local communities, we aspire to contribute to realizing a sustainable society.